\documentclass{INTERSPEECH2023}
\usepackage{multirow}
\interspeechcameraready

\title{Streaming Speech-to-Confusion Network Speech Recognition}
\name{Denis Filimonov, Prabhat Pandey, Ariya Rastrow, Ankur Gandhe, Andreas Stolcke}
\address{Amazon Alexa}
\email{\{denf,panprabh,arastrow,aggandhe,stolcke\}@amazon.com}

\begin{document}

\maketitle
 
\begin{abstract}
In interactive automatic speech recognition (ASR) systems, low-latency requirements limit the amount of search space that can be explored during decoding,
    particularly in end-to-end neural ASR.
In this paper, we present a novel streaming ASR architecture that outputs a confusion network while maintaining limited latency, as needed for interactive applications.
We show that 1-best results of our model are on par with a comparable RNN-T system, while the richer hypothesis set allows second-pass rescoring to achieve 10-20\% lower word error rate on the LibriSpeech task.
We also show that our model outperforms a strong RNN-T baseline on a far-field voice assistant task.
\end{abstract}

\section{Introduction}
\label{section:intro}

Prior to the arrival of end-to-end (E2E) neural models, most ASR systems were HMM-based, with acoustic models based on Gaussian mixture models (GMMs), and later, deep neural nets (DNNs).
These models generated an \emph{alignment} between a phonetic representation of the transcript and the sequence of acoustic frames.
E2E models, such as CTC, RNN-T, and LAS, largely dispensed with producing explicit alignments, 
    arguably to their advantage, because producing a fine-grained alignment is a difficult, error-prone, and often unnecessary task.
However, \emph{some} alignment is necessary for some purposes, for example, for a streaming ASR to know when an utterance starts and ends.
This need motivated recent work to achieve better end-of-utterance alignment in RNN-T systems~\cite{fast-emit-2021, unified-eos-2022}.
\cite{fast-emit-2021} reports not only reduced latency in end-of-sentence (EOS) detection, but also better accuracy by discouraging RNN-Ts from delaying the output too much.
We hypothesize that there may be some granularity of alignment between phone-level and utterance-level that would be most beneficial for accuracy and latency, depending on the task at hand.

Another challenge E2E systems face is the exploration of the hypothesis search space while decoding.
Because of the computational cost associated with maintaining each hypothesis, real-time streaming systems typically maintain only a handful of active hypotheses.
Moreover, because of recurrent dependencies, hypotheses form a prefix tree which limits diversity of the output, especially in long-form ASR (though there are some efforts to mitigate this problems \cite{Prabhavalkar2020LessIM, novak22grafting}).
Non-autoregressive (NAR) models \cite{Higuchi2020ImprovedMF,
streaming-nar-2021}, on the other hand,  approach the search problem very differently: they generate predictions for each unit (such as a word) independently, and then try to reconcile predictions in some way.
The main motivation for the NAR approach is computational efficiency, allowing everything to be computed in parallel,
    but their accuracy is not yet on par with autoregressive models such as RNN-Ts.
We hypothesize that one of the reasons may be that NAR methods produce too many hypotheses, making the task of finding the best one more difficult.
Additionally, NAR models are typically non-streaming (require the entire audio at once), though there are some streaming NAR models as well \cite{streaming-nar-2021}.

In this work, we propose an ASR model that strikes a balance between properties of regressive and non-autoregressive models.
Our model generates coarse, \emph{segment}-level alignment (each segment typically comprising one to three words).
Each segment is then decoded independently, as in NAR models, but decoding of each segment is autoregressive.
The output of our model is therefore a confusion network \cite{confusion-network}, each position-bin containing hypotheses for a single segment.
The confusion network is then rescored using a language model. 
Finally, each segment can be processed as soon as its final boundary is detected, thus making our model suitable for streaming ASR tasks.


\section{Related Work}
\label{section:related-work}

Transducer-based E2E ASR (RNN-T~\cite{graves2012sequence, rnnt2013} and related models, like
    Transformer-Transducer \cite{Zhang2020TransformerTA} and Conformer~\cite{conformer-2020})
have been popular choices for streaming ASR since their introduction.
Attention-based methods such as \cite{attention-asr-2015} and LAS~\cite{LAS2016},  on the other hand,
    have started as \emph{offline} methods that require the entire audio to be available at the start of recognition.
Subsequent work has relaxed this requirement by processing audio in fixed-size chunks \cite{mocha-2019, Dong2019SelfattentionAA, Tian2019SynchronousTF, triggered-attention-2020, Tsunoo2020StreamingTA, Miao2020TransformerBasedOC}.
\cite{scout-network-2020} introduced a separate "scout" network used to identify word boundaries instead of fixed-size chunks, giving reduced latency and improved accuracy.
Likewise, non-autoregressive models were initially introduced as offline methods \cite{Higuchi2020ImprovedMF}, though later work added a streaming capability \cite{streaming-nar-2021}.

Some papers have investigated improving representational power (in terms of the number of encoded hypotheses) of E2E ASR output by explicitly expanding the lattice \cite{Prabhavalkar2020LessIM, novak22grafting}, or by using a discrete, quantized state representation for the decoder \cite{Shi2022VQTRT}.
Rescoring of confusion network was investigated in \cite{iterative-decoding-2009}.

\section{Model Architecture}
\label{section:model}

In this section, we outline the general architecture of our model, while the specifics are described in Section~\ref{section:experiments}.

\begin{figure}[t]
  \centering
  \includegraphics[width=\linewidth]{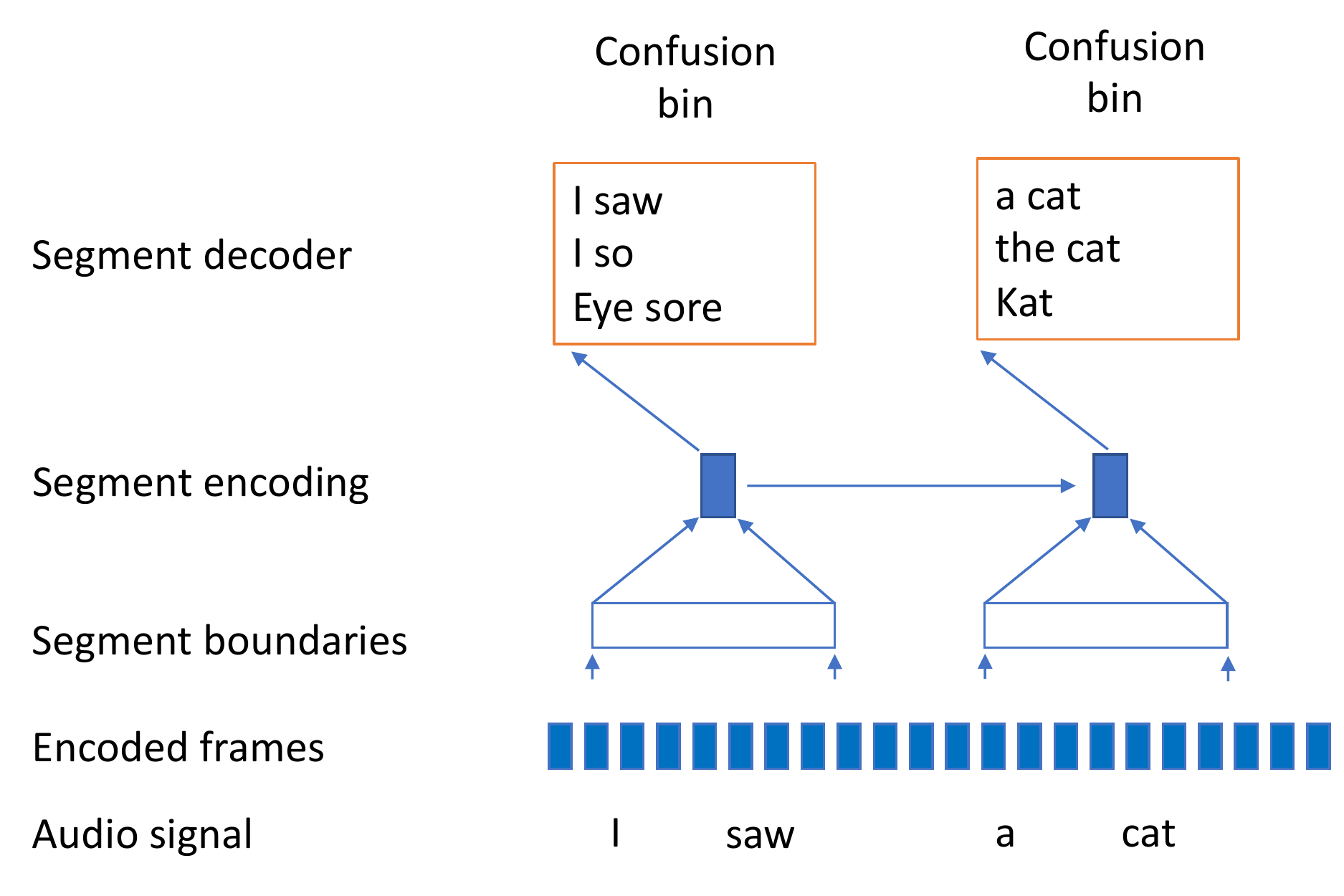}
  \caption{Schematic diagram of the speech-to-confusion betwork (S2CN) model.}
  \label{fig:s2cn-diagram}
\end{figure}

Figure~\ref{fig:s2cn-diagram} depicts the structure of the proposed model, comprising three main components:
audio encoder, segmenter, and a seq2seq model for mapping variable-length segments into variable-length word sequences.
The audio encoder (AE) maps input features, such as LFBE, to frame-level representations $h_t$. 
\[
h_t = AE(x_{\le t})
\]
There are no specific requirements on the encoder imposed by our model (indeed, we reuse the audio encoder structure from our RNN-T baseline), though, as our goal is steaming ASR, we limit ourselves to causal formulations.

The role of the segmenter is to identify continuous segments of one or more words.
We use BIO (begin-inside-outside) labels assigned to all frames, enabling us to have adjoining speech segments, as well as non-speech regions.
\[
s_t = S(h_{\le t+L}),~ s_t \in \{B, I, O\}
\]
In order to achieve high accuracy, the segmenter is allowed a lookahead of $L$ frames. 
The sequence of BIO labels produced by the segmenter can be transformed into a sequence of speech segments $ss_i = [start_i, end_i)$. 
We will use the subscript $i$ to index segments.

A number of seq2seq architectures can be employed to translate the sequence of frames $H_i = \{h_t : t \in ss_i\}$ into a sequence of labels $w_i^1 \ldots w_i^l$.
Taking into account that our segments are quite short (generally less than one second long), we propose the following architecture.
Our seq2seq model comprises three parts: a \emph{segment aggregator}, a \emph{segment encoder}, and a \emph{decoder}. 
The segment aggregator $\mathit{SA}$ encodes a variable-length frame sequence into a fixed-size representation.
In order to account for segmentation errors, we append a few frames on either side of the segment boundaries (pre- and post-roll).
Note that as long as the post-roll is within the segmenter's lookahead, it does not contribute to latency.
\[
\mathcal{H}_i = \mathit{SA}(\{h_t : t \in [start_i-\text{pre-roll}, end_i+\text{post-roll})\})
\]
The segment encoder is a recurrent model over the segments, its purpose is to add long-distance dependency into segment encodings more efficiently than could be done at the frame level.
\[
\hat{\mathcal{H}}_i = \mathit{SE}(\mathcal{H}_{\le i})
\]
Finally, the decoder is a recurrent (LSTM-based) auto-regressive model that generates the word-piece output $w_i^l$ from $\hat{\mathcal{H}}_i$, which is added to the LSTM's input after a linear projection.

For termination of the output we use an end-of-segment token, which is separate from end-of-sentence.
\[
w_{i,l} = D(\hat{\mathcal{H}}_i, s_{i, <l})
\]

The output of this model is a type of a confusion network in which each bin contains multi-word strings (segment-level n-best hypotheses), such as\\

\noindent
\begin{tabular}{|l|l|l|l|}
\hline
    bin 0 & bin 1 & bin 2 & bin 3 \\
\hline
    HANDED TO & ME & SAID THE CHIEF & \multirow{3}{*}{...} \\
    HAND IT TO & TO ME & SAYS THE CHIEF & \\
    HANDED & MY & SIGHED THE CHIEF & \\
\hline
\end{tabular}
\\

\noindent
Such a representation can encode a very large number of hypotheses and therefore
requires efficient rescoring methods, which we describe in the following section.

Perhaps one drawback of the confusion network (CN) representation is that it makes the use of external language models (LMs) in the first pass (e.g., via shallow fusion) more difficult.
Different strategies can be employed to bring an external LM into the first pass, for example, by adapting the LM to consume CN context, or by approximating the CN context by n-best, or by utilizing a deeper form of LM fusion \cite{jain2020contextual,sathyendra2022contextual}.
However, we leave the problem of first-pass biasing to future work and focus on second-pass rescoring.

\section{Rescoring Methods}
\label{section:rescoring}

We explore several strategies for rescoring confusion networks.
The simplest one is to extract n-best hypotheses based on their ASR score, and then rerank them based on a combination of ASR and LM scores.
Another simple method is to apply beam search, expanding one bin at a time, reranking hypotheses based on combined scores and retaining only top-n candidates.
This requires a language model that can score partial hypotheses (such as an autoregressive LM), but is suitable for streaming ASR: rescoring does not need to wait until the end of the utterance.
Finally, we can look at rescoring a confusion network as an instance of stochastic optimization.
The problem is to find the most likely hypothesis under the distribution defined by the combined ASR and LM weights: 
\[
\hat{W}_{1 \ldots k} = 
    \underset{W_{1 \ldots k}}{\mathrm{argmax}}~ \log p_\mathrm{lm}(W_{1 \ldots k}) + \alpha \sum_{i} \log p_\mathrm{asr}(W_i)
\]
where $p_\mathrm{lm}$ and $p_\mathrm{asr}$ are the LM and ASR scores, respectvely, and $W_i$ is a random variable that ranges over the string values in the $i$\textsuperscript{th} bin.
The argmax can be computed approximately using Gibbs sampling, similar to \cite{iterative-decoding-2009}.

\section{Experiments}
\label{section:experiments}

We now give more details of the model and present evaluation results on LibriSpeech and de-identified far-field voice assistant data.
Our LibriSpeech baseline is a streaming RNN-T with 7x1024 LSTM encoder layers with layer normalization~\cite{layernorm-2016} (LSTM-LN)  projected to 512, 2x512 LSTM-LN decoder layers, and a 2.5k word-piece output vocabulary.
For the speech-to-confusion network (S2CN) model, we use the same encoder structure, except we reduce the number of layers to 6 in order to keep the overall number of parameters on par with the baseline.
The inputs are 3 stacked 64-dimensional LFBE features; each input frame thus represents 30ms of audio.
To train the S2CN model, we use forced-alignment data.\footnote{https://github.com/CorentinJ/librispeech-alignments}
All models were trained on 24 V100 GPUs for 150k steps using an Adam optimizer.

\begin{figure*}[ht]
  \centering
  \includegraphics[width=1.0\textwidth]{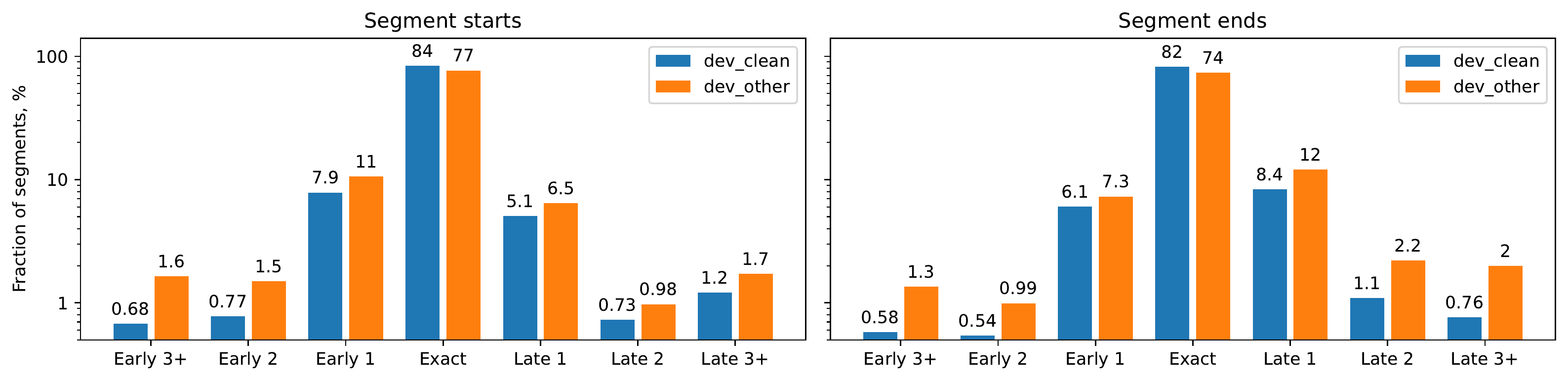}
  \caption{Segment start/end predictions relative to the closest word boundary. In the left plot, "Early 2" indicates the percentage of segments that start two frames earlier than the closest word start boundary.}
  \label{fig:segmenter-accuracy}
\end{figure*}

\textbf{Audio segmenter.} 
We observe that word-level segmentation is challenging; therefore we relax the problem by allowing a segment to contain multiple words.
The motivation is similar to the ``scout network'' of Wang~et~al.~\cite{scout-network-2020}. However, there are a number of important differences: we use BIO encoding, which allows us to detect and skip non-speech segments. Instead of training a separate large network just for segmentation based on LFBE input we use a small model, but pass it the output of the audio encoder as depicted in Figure~\ref{fig:s2cn-diagram}.
Specifically, we use transformer layers with a fixed-size window (16 past and 2 future frames) with 2 heads and attention model size (d\_model) of 256  and feed-forward model size (d\_ff) of 512.
This model is trained to predict word-level boundaries (using the same BIO coding and cross-entropy loss), with an added  heuristic that decides which word boundaries should be segment boundaries based on the confidence and the length of the current segment.
A segment spans 2.2 words on average.

\begin{figure}[ht]
  \centering
  \includegraphics[width=0.9\linewidth]{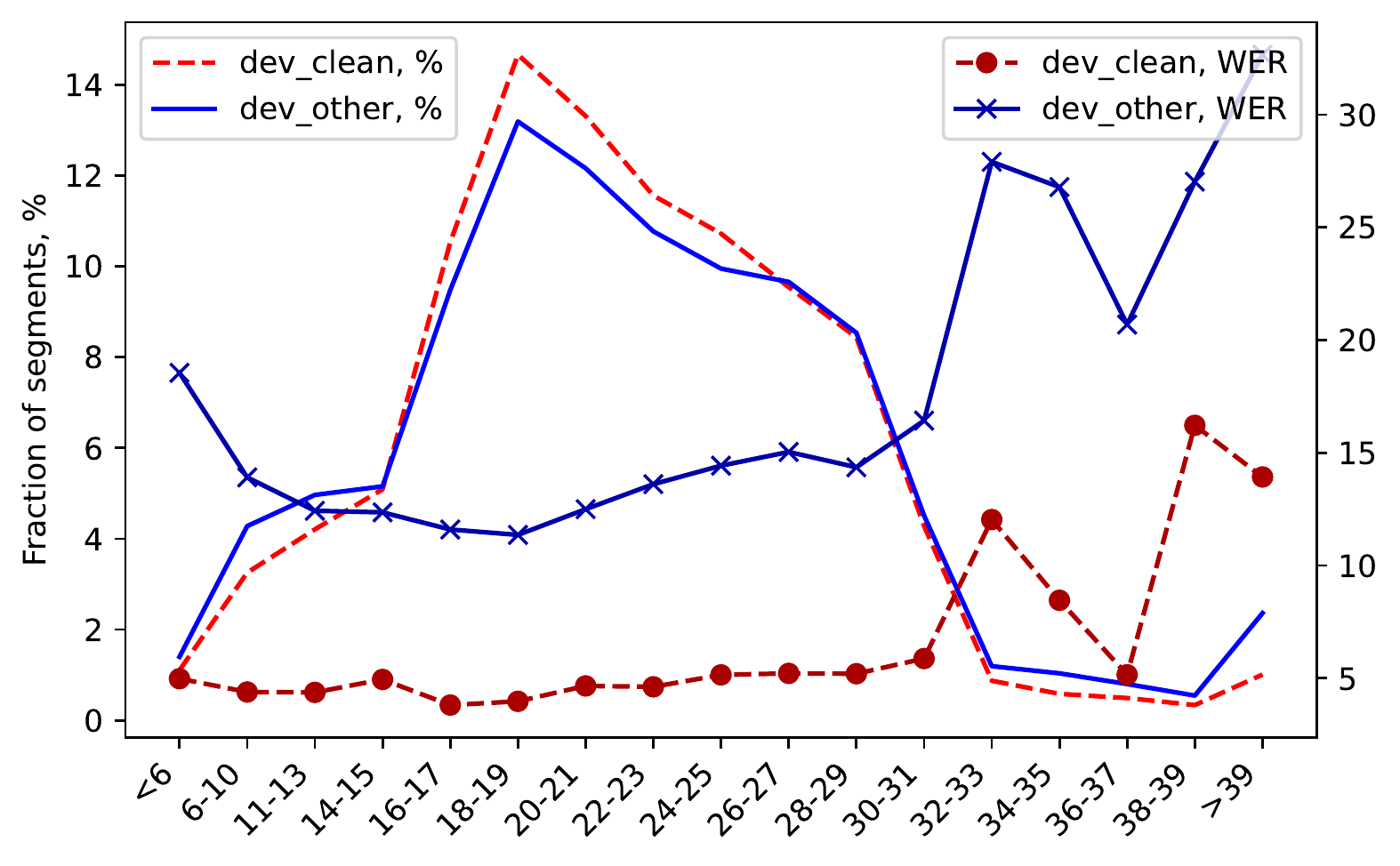}
  \caption{Segment distribution and accuracy. The horizontal axis denotes the range of segment lengths in 30ms frames.}
  \label{fig:segment-dist}
\end{figure}

Figure~\ref{fig:segmenter-accuracy} shows the accuracy of segment boundaries predictions relative to the closest word boundary aligned to that segment. Note that not only are we able to achieve high accuracy even in noisy conditions, but the majority of errors are only one frame off.
In Figure~\ref{fig:segment-dist}, we show the distribution of segment lengths and WER.
The majority of segments are between 16 and 30 frames (0.5-0.9 seconds) long, and the WER for longer segments rises sharply.
This degradation may occur because long segments are underrepresented in training and are acoustically more challenging
    (and therefore harder to break into shorter segments).

The \textbf{segment aggregator} constructs a fixed-size representation of a segment from the output of the audio encoder and the segment boundaries.
The simplest aggregator computes a mean of the encoded frames. However, we opt for a multihead attention layer with 4x256 heads, the aforementioned mean as the query, and the encoded frames as the keys and values.
We also add logits from the segmenter output to inform the aggregator about likely word boundaries within the segment.
Additionally, owing to the fact that most segmentation errors are just a frame off, we append two frames on either side of the segment (pre- and post-roll) to the aggregator inputs.

The \textbf{segment encoder} gives the model the opportunity to incorporate long-distance dependencies between segments, rather than at the frame level.
We use a single 512 LSTM-LN layer.

The \textbf{decoder} is a 2x512 LSTM-LN model with the encoded segment projected to 384 dimensions and appended to each layer of the decoder's LSTM inputs.

During training, first segment boundaries are produced by the segmenter, then transcription is aligned to the segments based on word and segment boundaries, so that each word belongs to one segment, then each segment is trained to produce its portion of the transcript ending with a special end-of-segment token.
We use multi-task training with cross-entropy losses for the segmenter labels and the transcription.
Because the segmenter's output is initially random, we use ``teacher'' segmentations derived from true word boundaries.
This teacher segmentation is then phased out exponentially as the training progresses.

\begin{table}[tb]
    \caption{1-best and oracle (in parentheses) WER on LibriSpeech. For RNN-T, oracle numbers are out of 8-best hypotheses, for S2CN, the oracle WER is computed on confusion networks with 8-best hypotheses per bin.}
    \label{tab:libri-ablation}
    \centering
    \begin{tabular}{l|c|c|c}
    \hline
        Model & Params & dev\_clean & dev\_other\\
    \hline
        RNN-T & 63.0M & 5.21 (2.90) & 14.4 (10.5) \\
    \hline
        S2CN & 62.7M & 4.92 (1.30) & 14.4 (5.89) \\
        -segment aggregator (A) & 59.5M & 5.37 (1.49) & 14.6 (6.22) \\
        -segmenter logits (B) & 62.7M & 5.36 (1.39) & 14.1 (5.52) \\
        -pre-/post-roll (C) & 62.7M & 5.51 (1.45) & 14.6 (5.84) \\
        -segment encoder (D) & 60.6M & 5.14 (1.35) & 14.5 (5.84) \\
    \hline
    \end{tabular}
\end{table}

Table~\ref{tab:libri-ablation} compares our model against the RNN-T baseline.
Looking at 1-best result, S2CN performs better than the baseline on \texttt{dev\_clean} data and similarly on the noisier \texttt{dev\_other} set.
More importantly, the oracle WER (in parentheses) is nearly half that of the RNN-T's, owing to the factorial representation in a confusion network.
Table~\ref{tab:libri-ablation} also shows results from several ablation studies:
in (A) we replace the multihead attention segment aggregator with a simple mean;
in (B) we withhold segmenter logits from the segment aggregator input; 
in (C) we omit the two-frame pre-/post-roll;
and in (D) we remove the segment encoder (the output of the segment aggregator goes directly to the decoder).
Notably, none of the components seem to be critical for the system, 
    with perhaps pre-/post-roll being the most important one.
    This confirms our intuition that missing a part of a word is more damaging to the performance than including a word fragment from an adjacent segment.
Interestingly, the omission of the segment encoder has a smaller impact than we might have expected, 
    which may indicate that even in long LibriSpeech utterances long-distance dependencies do not seem to be utilized well by the first pass.

\begin{table}[tb]
    \caption{N-best rescoring results on LibriSpeech (WER). \#hyp denotes the number of hypotheses.}
    \label{tab:n-best-rescoring-libri}
\centering
    \begin{tabular}{|l|c|c|c|c|c|}
        \hline
        \multicolumn{2}{|c|}{~} & \multicolumn{2}{c|}{clean} & \multicolumn{2}{c|}{other} \\
        \hline
        Model & \#hyp & dev & test & dev & test \\
        \hline \hline
         RNN-T & 1 & 5.21 & 5.68 & 14.4 & 14.3 \\
         n-best & 8 & 4.03 & 4.62 & 12.3 & 12.6 \\
        \hline
        S2CN   &  1  & 4.92 & 5.18 & 14.4 & 14.6 \\
        n-best &  8  & 3.66 & 3.88 & 12.1 & 12.4 \\
        n-best & 20  & 3.45 & 3.72 & 11.5 & 11.9 \\
        n-best & 100 & 3.32 & 3.63 & 10.9 & 11.5 \\
        n-best & 1000 & 3.23 & 3.57 & 10.4 & 11.1 \\
        \hline
    \end{tabular}
\end{table}

In Table~\ref{tab:n-best-rescoring-libri}, we present n-best rescoring results using an autoregressive language model trained on the LibriSpeech text corpus.
The model consists of 6 transformer blocks (8 heads, d\_model=512, d\_ff=2048).
We use a beam of 8 for decoding in both RNN-T and S2CN; therefore, RNN-T is limited to 8-best output, while we can extract much deeper n-best lists from confusion networks.
Interestingly, even if we limit n-best lists generated from confusion networks to 8, we gain more from rescoring via S2CN.
We attribute this to better diversity of the n-best lists extracted from confusion networks---n-best lists from RNN-T are prefix trees by construction, therefore most of the variation tends to occur at the end.

\begin{table}[tb]
    \caption{Confusion network rescoring results on LibriSpeech. \#hyp denotes the average number of hypotheses.}
    \label{tab:cn-rescoring-libri}
    \centering
    \begin{tabular}{|l|c|c|c|c|}
    \hline
            & \multicolumn{2}{c|}{test\_clean} & \multicolumn{2}{c|}{test\_other} \\
    \cline{2-5}
        Method   & WER & \#hyp & WER & \#hyp \\
    \hline \hline
        100-best       & 3.63 & 100  & 11.5 & 100 \\
   Gibbs (L2R), 1-pass & 3.55 & 10.4 & 10.9 & 19.9 \\
   Gibbs (L2R), 3-pass & 3.54 & 30.9 & 10.8 & 59.8 \\
   Gibbs (H2L), 1-pass & 3.57 & 10.4 & 10.9 & 19.9 \\
   Gibbs (H2L), 3-pass & 3.55 & 30.9 & 10.8 & 59.8 \\
        Streaming      & 3.59 & 9.35 & 10.9 & 18.6 \\
    \hline
    \end{tabular}
\end{table}

In Table~\ref{tab:cn-rescoring-libri}, we evaluate the different rescoring methods described in Section~\ref{section:rescoring}.
Prior to rescoring, we prune confusion bins by removing hypotheses with less than $0.005$ probability mass.
In many bins this process removes all but one hypothesis, substantially reducing the number of hypotheses to score.
With the Gibbs sampling method, we evaluate two strategies for selecting in which order bins (variables) are explored: either left-to-right (L2R) or in the order of internal bin entropy, high-to-low (H2L).
We also evaluate how quickly the algorithm converges by taking one or three passes through all bins.
We find that there is no significant difference between the two bin traversal orders and that there is very little room for further improvement after a single pass through the bins.
Finally, we observe that streaming rescoring performs only slightly worse than Gibbs rescoring, and that all confusion network rescoring methods significantly outperform n-best rescoring both in terms of WER and the number of hypotheses scored.

Finally, we evaluate our model on a far-field voice assistant task.
The inputs are 3 stacked 64-dimensional LFBE features.
The baseline RNN-T model consists of 2x1280 FLSTM feature encoder followed by 2x time reduction and then by 6x1280 LSTM-LN layers.
The decoder is 2x1280 LSTM with 4k word-piece output.
The encoder is pretrained using a cross-entropy objective on a large amount of de-identified forced-aligned transcriptions (human- and machine-made).
The S2CN model uses the same pretrained encoder, except its output is projected to 768 dimensions.
The rest of the S2CN model is the same as in the LibriSpeech experiment, except that the segment encoder and decoder use 768-dimensional LSTM-LN layers
    and the audio segmenter window is reduced to 8 past and 2 future frames, since in this setup each frame is twice as long (60ms).
Both RNN-T and S2CN models are similar in size (148M and 149.5M parameters, respectively) and are trained on the same data on 8 V100 GPUs for 600k steps using an Adam optimizer.
As in the LibriSpeech experiments, we use a beam size of 8 for inference.

\begin{table}[tb]
    \caption{WER reduction relative to RNN-T 1-best (in percent). Positive numbers represent lower WER.}
    \label{tab:alexa-eval}
    \centering
    \begin{tabular}{|l|c|c|c|c|}
        \hline
             & \multicolumn{2}{c|}{RNN-T} & \multicolumn{2}{c|}{S2CN} \\
        \cline{2-5}
        Dataset & 1-best & n-best & 1-best & n-best \\
        \hline \hline
        General & - & 2.8 & -0.1 & 6.32 \\
        \hline
     Rare words & - & 4.1 & -2.2 & 7.02 \\
        \hline
   3rd-party apps & - & 0.8 & 0.0 & 4.72 \\
        \hline
      Messaging & - & 1.9 & 0.8 & 4.26 \\
        \hline
    \end{tabular}
\end{table}

In Table~\ref{tab:alexa-eval}, we present the results (1-best and rescoring) on four datasets: a random sample of de-identified voice assistant requests,
a subset of queries containing at least one rare\footnote{Words occurring less than 5 times in a sample of 1.6M utterances.} word,
a set of queries to 3rd-party applications, and requests to the messaging application.
For rescoring, we use a transformer-based LM comprising 6 transformer blocks (8 heads, d\_model=512, d\_ff=2048) that was trained on a large set of in-domain data.
We only present n-best rescoring results in this case because most requests are short (approximately 5 words on average, spanning 1 to 3 bins), and exhaustive enumeration is therefore feasible. We limit the number of S2CN n-best to 100, though many utterances have fewer hypotheses.
We observe that 1-best performance of S2CN is comparable to RNN-T, but the richer output enables greater improvement in second-pass rescoring. 

\section{Conclusions}
\label{section:conclusions}

We have presented a novel ASR architecture that learns to produce alignments for multi-word utterance segments, along with confusion network output, trading some long-distance dependencies for richer local representations of the hypothesis space.
We showed that this model has 1-best performance comparable to RNN-T, but its more expressive output enables the second-pass rescoring to achieve over 20\% lower WER on LibriSpeech \texttt{test\_clean} and over 10\% on \texttt{test\_other}.
We have also evaluated several rescoring methods for confusion networks that proved significantly more efficient than n-best rescoring.
Finally, we showed that our model also yields gains over RNN-T-based n-best rescoring on a far-field voice assistant task.

\bibliographystyle{IEEEtran}
\bibliography{s2cn}

\end{document}